\documentclass[%
 reprint,
superscriptaddress,
 amsmath,amssymb,
 aps,
 prl,
]{revtex4-2}

\usepackage{graphicx}
\usepackage{dcolumn}
\usepackage{bm}
\usepackage{xcolor}

\begin{document}


\title{Microswimmers create bicontinuous emulsions in binary fluids}


\author{Harinadha Gidituri}
\email{g.harinadha@che.iitkgp.ac.in} 
\affiliation{Department of Chemical Engineering, Indian Institute of Technology Kharagpur - Kharagpur 721302,
West Bengal, India
}
 \author{Sotiris Samatas}
\affiliation{EMBL Barcelona, Barcelona Biomedical Research Park, 08003, Barcelona, Spain 
}
\affiliation{Barcelona Collaboratorium for Modelling and Predictive Biology, 08005 Barcelona, Spain 
}

\author{Juho S. Lintuvuori}
\email{juho.lintuvuori@u-bordeaux.fr}
\affiliation{Univ. Bordeaux, CNRS, LOMA, UMR 5798, F-33405 Talence, France 
}%


\date{\today}

\begin{abstract}

We consider a generic case of neutrally wetting microswimmers in symmetric mixtures of two phase separating fluids, using hydrodynamic simulations. The swimmers spontaneously emulsify the two fluids into bicontinuous foam-like state.   
The two principal activity components: source dipole (self-propulsion) and force dipole (active mixing),  create a twofold mechanism to stabilise the structures. When the self-propulsion is too strong, 
the swimmers cross the interfaces rapidly and the two fluids will phase separate. Below this threshold, the active stresses from the force dipoles, stabilise a dynamic and bicontinuous foam-like state. When the activity is turned off, the system relaxes into a kinetically trapped bicontinuous state, with particles permanently trapped at the interfaces. Our results provide a microscopic route to tunable  active emulsions, with implications for bacterial suspensions and synthetic active matter.
\end{abstract}

\maketitle

 {\it Introduction.--}
Liquid-liquid phase separation is an equilibrium process, where an initially homogeneous mixture of two immiscible fluids, demixes into domains separated by interfaces~\cite{Kendon_JFM_2001}.  In a symmetric binary mixture, the phase separation occurs via spinodal decomposition where the domains are interconnected and bicontinuous. 
The full phase separation can be arrested for example by adding particles (colloids or nanoparticles), leading to the formation   
of kinetically arrested structures, such as bicontinuous jammed emulsions (bijels)~\cite{Stratford_Science_2005,Cates_SM_2008}. 
An alternative way to suppress the phase separated state is by injecting mixing energy into the system, for example by turbulence~\cite{Perlekar_POF_2012,Perlekar_PRL_2014,Dennis_PRL_2021,Perlekar_SR_2017}, or internally using active  fluids~\cite{Adkins_Science_2022,Fernando_PRL_2022,Tayar_NatMat_2023,Tiribocchi_PRL_2015,cates2025active,diaz2025activity}. 
  
Continuum theories and  simulations have predicted arrested~\cite{Tiribocchi_PRL_2015,cates2025active,singh2019hydrodynamically} and suppressed phase separation~\cite{Fernando_PRL_2022} by active stresses.  
Experiments with phase separating binary mixtures of active and passive fluids, show the formation of an emulsion like state at high activities~\cite{Adkins_Science_2022}, and control of the liquid-liquid phase separation behaviour~\cite{Tayar_NatMat_2023}. Active Brownian particles (ABPs) with an affinity to one of the fluids, have showed the formation of foams due to an active pressure at the interfaces~\cite{diaz2025activity}, while recent experiments demonstrate the formation of bicontinous emulsions in, close to symmetric, mixtures of passive and active fluids~\cite{gulati2026bicontinuity}.

The activity can also be injected microscopically adding bacterial swimmers in passive phase separating binary mixtures~\cite{cheon2024motile,cheon2025motility}. Now both the particle activity (swimming stresses) and wetting induced interfacial trapping forces play a role~\cite{mishra2026interface}, and can lead to the formation of bacteria based Pickering emulsions~\cite{Ali_TB_2025} and bacterial degradation of oil spill droplets~\cite{Prasad_Science_2023}.   
Very recent experimental studies~\cite{Yang_Arxiv_2025,Chang_Arxiv_2025} on asymmetric mixtures of dextran (DEX) and polyethylene glycol (PEG) in the presence of motile bacteria have shown arrest of droplet coarsening, and revealed a rich variety of emergent phases: (i) self-spinning droplets, (ii) droplet chains, and (iii) capillary bridges populated by bacterial clusters~\cite{Yang_Arxiv_2025,Chang_Arxiv_2025}. 
Further experiments show, that wetting induced trapping at the interfaces coupled with swimming stresses can lead to active bacterial adhesion~\cite{Yang_Arxiv_2026} and allow long range cell capture~\cite{Wang_bioarxiv_2026}. 
However, it is currently not known if the hydrodynamics of bacterial or artificial microswimmers  
can create bicontinuous emulsions or swimmer stabilised foams. 

 Using hydrodynamic simulations we study the effects of neutrally wetting microswimmers to the phase morphology  of a symmetric mixture of passive phase separating binary fluids. Starting from an initial fully phase separated initial state, we observe spontaneous formation of bicontinuous emulsions due to hydrodynamic mixing by the swimmers. The physics can be captured considering two capillary numbers: $\mathrm{Ca}_{B_{1,2}} = \tfrac{\eta | B_{1,2}|}{\gamma}$, which measure the ratio between active hydrodynamic flows arising from self-propulsion $B_1$ (source dipole) or active mixing $B_2$ (force dipole) flows, and the interfacial tension $\gamma$.   We show that the structures are stabilised by force dipole flows for sufficiently high $\mathrm{Ca}_{B_2}$. The mechanism is likely similar to what is observed in continuum~\cite{Tiribocchi_PRL_2015,gulati2026bicontinuity}, however, in our case the swimmers have a finite size and are self-propelled.  
 Increasing the swimming speed $u_0\sim B_1$ leads to the destabilisation of the foam state, and a full phase separation of the two fluids is observed. Our results hold for both pusher (extensile) and puller (contractile) swimmers for various volume fractions. Finally, we show that when the activity is turned off, the system evolves into a kinetically trapped state akin to bijels observed with passive colloids in binary fluids~\cite{Stratford_Science_2005}.  
 
 {\it Computational model and parameters.--}
We model the microswimmers as spherical squirmers with radius $R$~\cite{Juho_SM_2016,Zaiyi_EPJE_2018}, where a surface slip boundary condition is implemented at  the particle surface~\cite{Pagonabarraga_JNFM_2010,magar2003nutrient}
\begin{equation}
 v_{s} = B_{1} \sin\theta + B_{2} \sin\theta \cos\theta .
\label{eq:slip velocity}
\end{equation}
The first term corresponds to a source dipole, it is responsible for the bulk swimming speed $U_0 = \frac{2}{3}B_1$. The second term corresponds a force dipole (stresslet) and it contributes to the mixing of fluid around the squirmer~\cite{Lin_JFM_2011}. The ratio of their amplitudes defines a squirming parameter $\beta=B_2/B_1$, which characterises the swimmer type: pullers (pushers) correspond to $\beta > 0$ ($\beta < 0$) while $\beta = 0~(\pm \infty)$ is a neutral (shaker) swimmer. 

The binary fluid mixture is modelled by using Ginzburg-Landau free energy functional~\cite{Kendon_JFM_2001},
 \begin{equation}\label{eq:free_energy}
 F[c] = \int dV \left({-\frac{A}{2}c^2 + \frac{B}{4}c^4 + \frac{\kappa}{2} |\nabla c|^2} \right)
 \end{equation}
where $-A = B >0$ and surface penalty $\kappa$ are constants, and $c$ is the phase composition, where $c^*=\pm 1$ are the equilibrium compositions of the two fluids.
We use $-A=B=0.00258$ and $\kappa=0.004$, which give an interfacial width~\cite{Kendon_JFM_2001} $\chi_{0} = \sqrt{2 \kappa/|A|} \approx 1.76$ and interfacial tension $\gamma= \sqrt{8 \kappa |\mathcal{A}|^3/9 \mathcal{B}^2}\approx 0.003$ in simulation units. 
The temporal evolution of the phase field variable $c$ is governed by a Cahn-Hilliard advection-diffusion equation,
\begin{equation}\label{eq:Arrhenius}
\partial_{t} c + {\bf{v}}\cdot \nabla c = M \nabla^2 \mu,
\end{equation}
where ${\bf{v}}$ is the fluid velocity, $M=5$ is the mobility and $\mu$ is the chemical potential derived from the free energy functional in Eq. \ref{eq:free_energy} by $\mu = \frac{\delta G}{\delta c} = A c + B c^3 - \kappa |\nabla^2 c|$.
The coupled hydrodynamic equations are solved using a hybrid finite difference lattice Boltzmann (LB) scheme~\cite{Kendon_JFM_2001,Hari_PoF_2021,Gidituri_PRF_2022} using the open source LB code Ludwig~\cite{kevinstratford_2024_12822477,ludwigcode}. 

We consider finite size particles with $R=6.0$ in a density matched solution $\rho = 1$, where both the fluids have a viscosity $\eta=0.625$ in simulation units, in a cubic 
$26.7R\times 26.7R\times 26.7R$
simulation box with periodic boundary conditions. The particles have equal affinity to both fluids, which is typical in bacterial experiments~\cite{cheon2024motile,Yang_Arxiv_2026}. The dynamic state is given by Reynolds number $\mathrm{Re}= \frac{\rho R u_{0}}{\eta}\lesssim 10^{-2}$ which disregards inertial effects. To allow comparison with experiments, we describe the results considering two dimensionless capillary numbers: $\mathrm{Ca}_{B_{1,2}} = \tfrac{\eta |B_{1,2}|}{\gamma}$, which give the ratios between hydrodynamic forces (source and force dipoles) and interfacial forces.

The phase separation process is characterized by calculating the time evolution of a typical domain length-scale $L(t)$ 
defined as the inverse of the first moment of the spherically averaged structure factor, $S(k, t )$ 

\begin{equation}
    L(t) = 2\pi\frac{\int S(k,t) dk}{\int k S(k,t) dk}
\end{equation}

where $k = |k|$ is the modulus of the wave vector in Fourier space, and

\begin{equation}
    S(k,t) = <c({\bf{k}},t)c({\bf{-k}},t)>
\end{equation}

with $c({\bf{k}},t)$ the spatial Fourier transform of the order parameter $c$. The angle brackets denote an average over a shell in ${\bf{k}}$ space at fixed $k$.
 
\begin{figure}[tb]
\includegraphics[width=1.0\columnwidth]{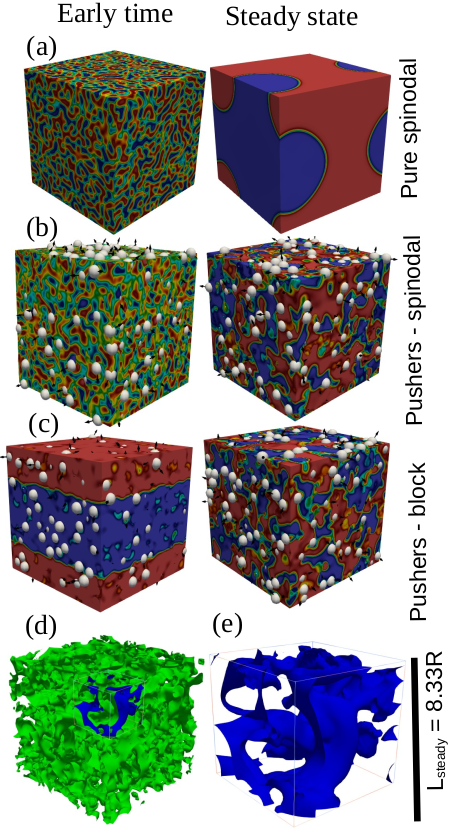}
\caption{The time evolution of the composition $c$ of the two fluids (left panel; early times, right panel; steady state). (a) Pure fluid with a fully mixed initial state, (b) pushers with fully mixed initial state, (c) pushers with phase separated initial state. (d,e) interface profile at the steady state with (d) the whole simulation box and (e) the characteristic length-scale of the foam-like structure. For (b-e) the data correspond to swimmer volume fraction $\phi \approx 0.2$, with $\mathrm{Ca}_{B_1} \approx 3.0$ and $\mathrm{Ca}_{B_2} \approx 7.0$. The color code corresponds to $c$ with red (blue) corresponding to $c=+1$ ($c=-1$). The white spheres are the swimmers.} 
\label{fig:SNapshots_pure_passive_active_comparison}
\end{figure}

\begin{figure}[tb]
\includegraphics[width=1.0\columnwidth]{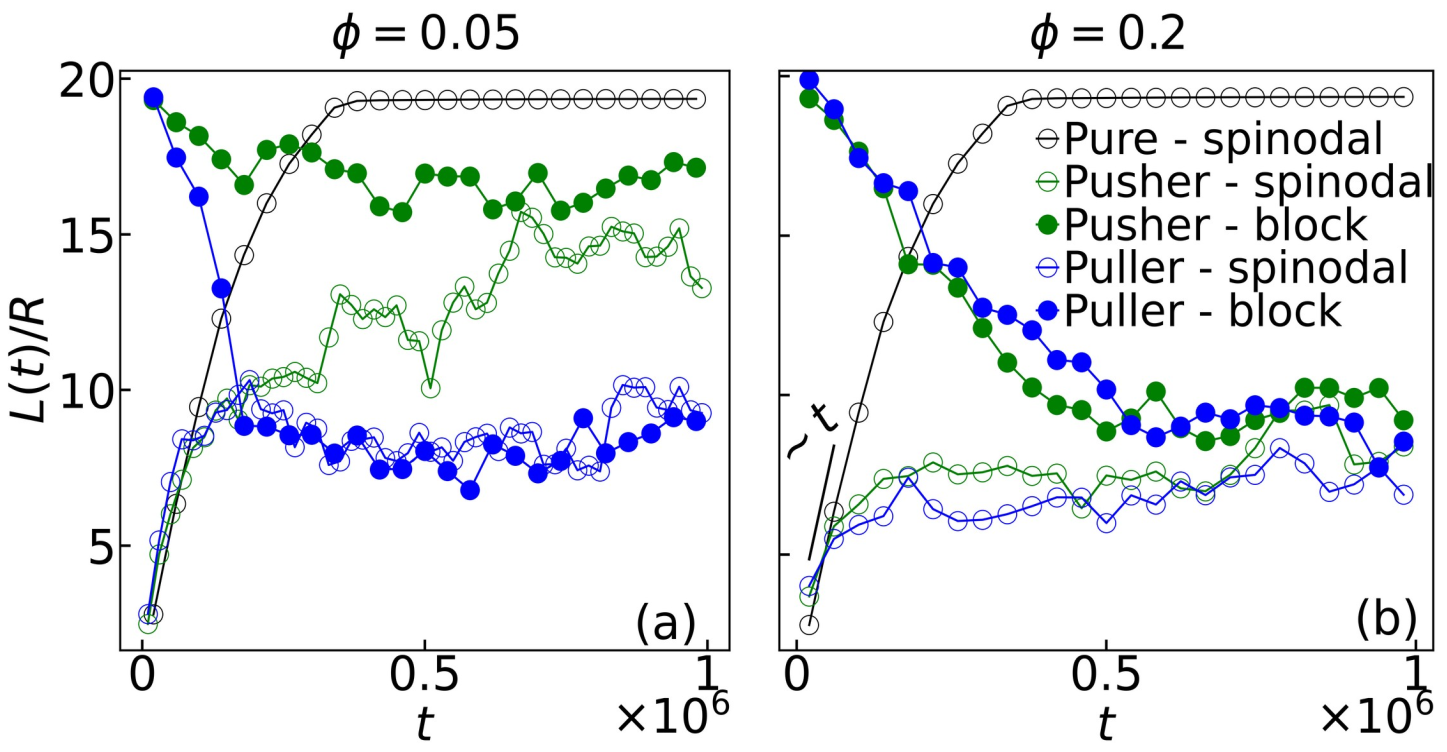}
\caption{The time evolution of the domain size $L(t)$ for pushers (green symbols) and pullers (blue symbols) starting from fully phase separated (full symbols) and fully mixed (open symbols) states. (a) Swimmer volume fraction $\phi \approx 0.05$, $\beta \approx \pm 5.0$ corresponding to $\mathrm{Ca}_{B_1} \approx 2.0$ and $\mathrm{Ca}_{B_2} \approx 10.0$. (b) $\phi \approx 0.2$, $\beta \approx \pm 2.33$ corresponding to $\mathrm{Ca}_{B_1} \approx 3.0$ and $\mathrm{Ca}_{B_2} \approx 7.0$. The black curve correspond to the quench of pure binary fluid.} 
\label{fig:domain_pure_passive_active_comparison}
\end{figure}

\begin{figure}[tb]
\includegraphics[width=0.485\columnwidth]{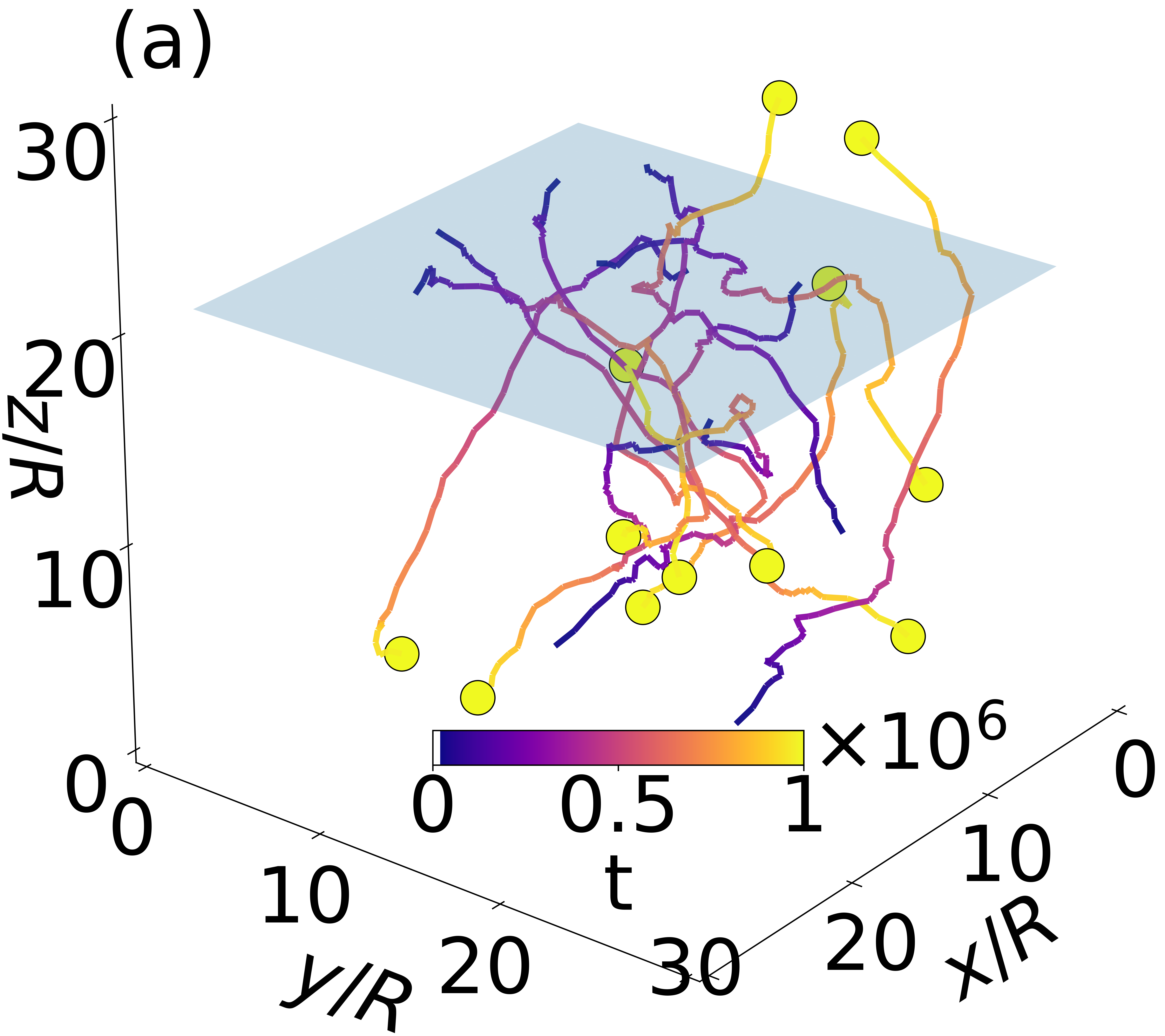}
\includegraphics[width=0.5\columnwidth]{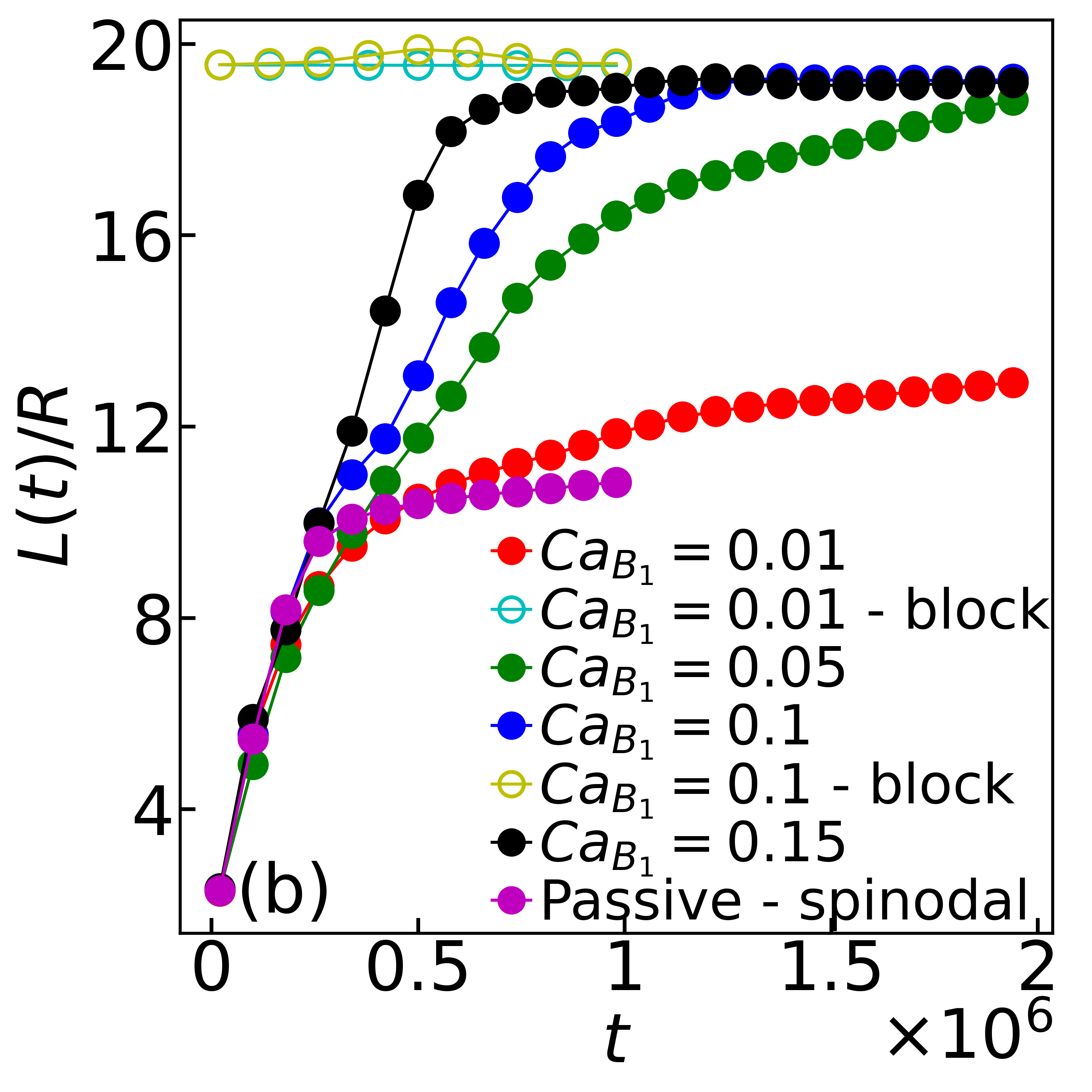}

\caption{(a) Observed trajectories of 12 selected neutral squirmers in a fully phase separated case. The data corresponds to $\mathrm{Ca}_{B_1} \approx 0.1$ and $\mathrm{Ca}_{B_2}$=0. The grey surface shows the interface. (b) Time evolution of the domain length-scale $L(t)$ for neutral swimmers at $\phi \approx 0.2$ for various $\mathrm{Ca}_{B_1}$.} 
\label{fig:neutral_CaB1_0p1}
\end{figure}

\begin{figure}[tb]

\includegraphics[width=1.0\columnwidth]{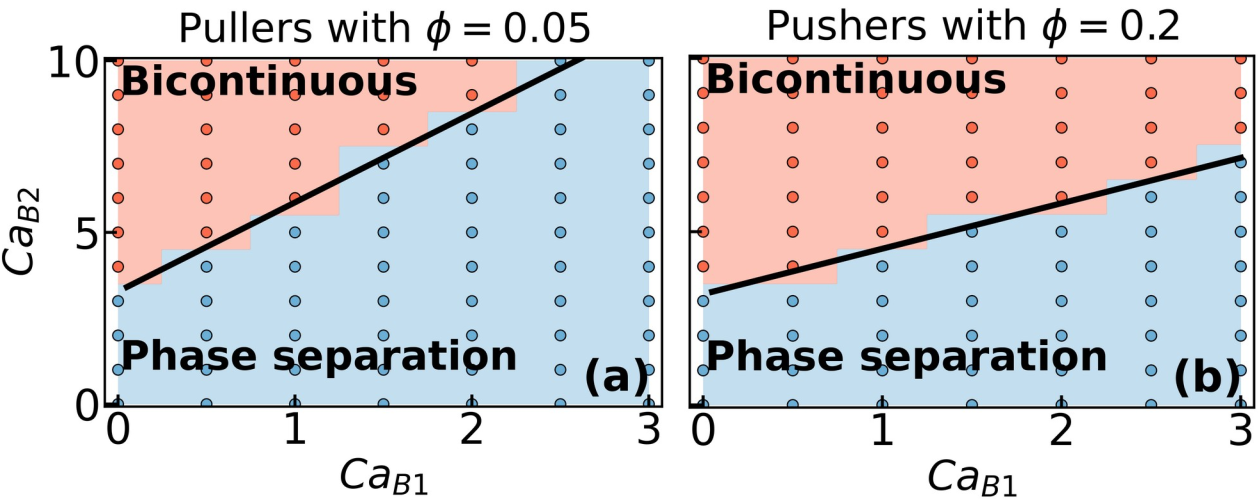}
\caption{A steady state phase diagram  between bicontinuous foam-like state (red symbols) and full phase separation (blue symbols) for (a) pullers with $\phi\approx 0.05$ and (b) pushers with $\phi\approx 0.2$. The solid lines correspond to a linear fit $B^*_2\approx AB_1 +C$ (see text for details).} 
\label{fig:phase_diagram}
\end{figure}

\begin{figure}[htb!]
\includegraphics[width=1.0\columnwidth]{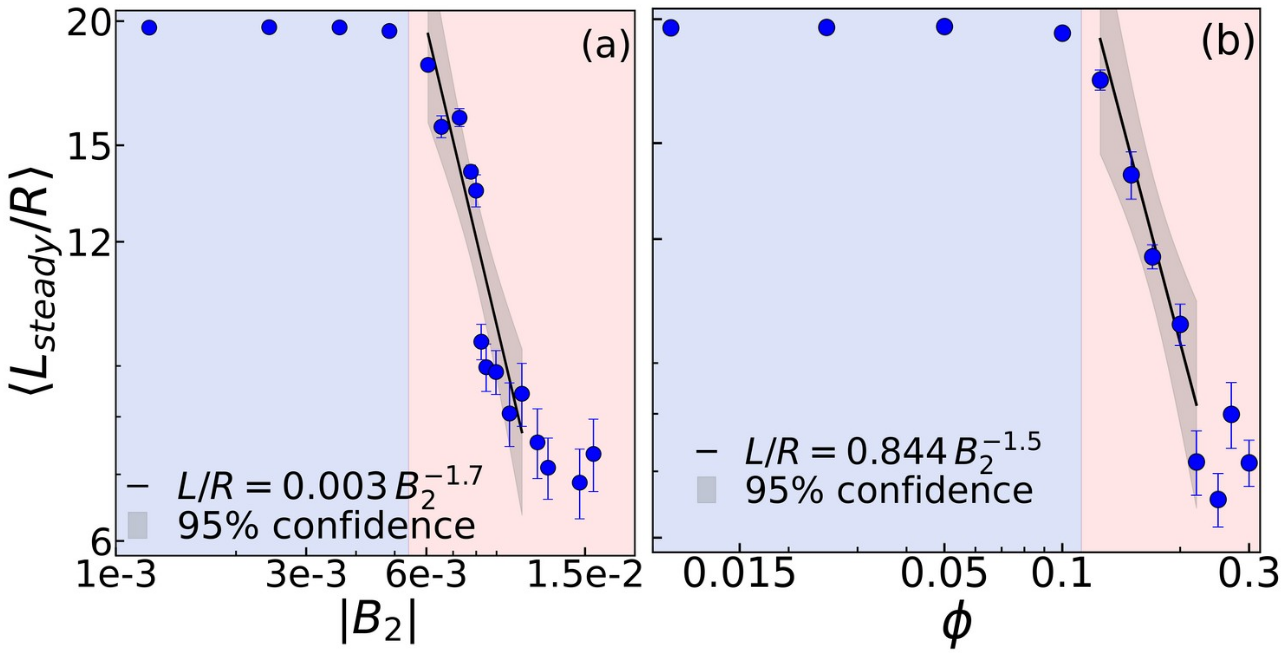}

\caption{ 
The average steady state domain lengths-scale $L$ for pushers as a function of (a) force dipole strength $B_2$ with $\phi\approx 0.2$ and (b) swimmer volume fraction $\phi$ with $B_2\approx 0.03391$ ($\mathrm{Ca}_{B_2}\approx 7.0$). The data corresponds to $\mathrm{Ca}_{B_1}\approx 3.0$ and the red (blue) shading marks the emulsion (phase separated) state. } 
\label{fig:B2_and_Phi_effect}
\end{figure}

\begin{figure}[tb]
\includegraphics[width=1.0\columnwidth]{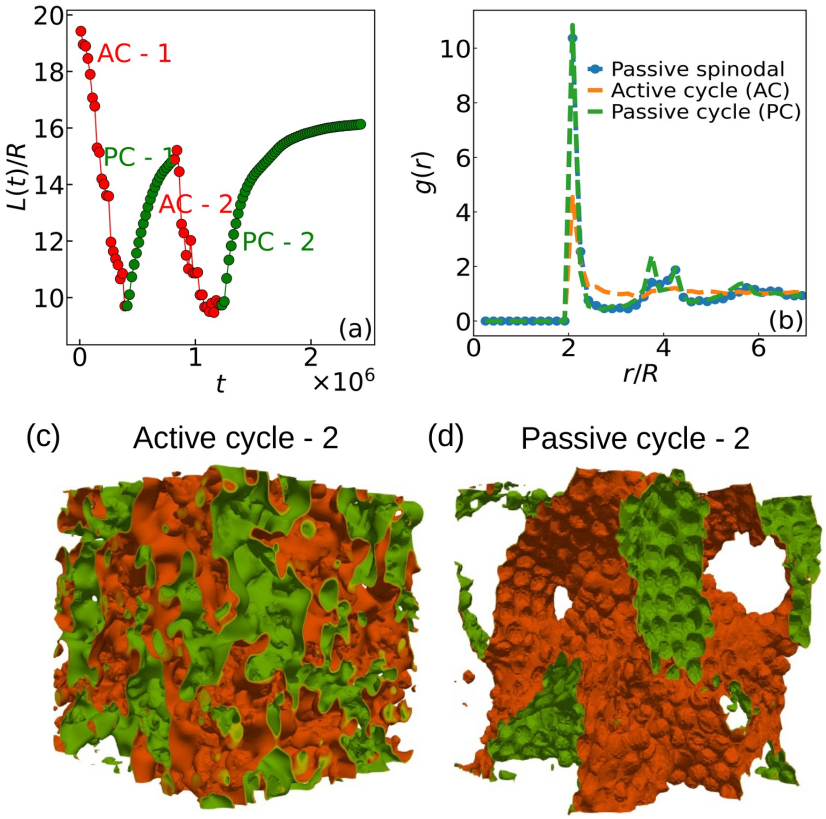}
\caption{Kinetically trapped gels by activity cycling. (a) $L(t)$ for active (red) and passive (green) cycles. (b) Pair distribution function $g(r)$ at the end of active (orange) passive (green) cycles. The blue curve corresponds to $g(r)$ of quenched passive colloids.  Comparison of radial distribution function for active and passive cycles against passive.  The interface morphology visualised at the end of (c) active cycle and (d) passive cycle. The colloids are omitted for clarity. The data correspond to pushers ($\beta\approx -2.33$) with $\mathrm{Ca}_{B_1}\approx 3.0$, $\mathrm{Ca}_{B_2}\approx 7.0$ and $\phi\approx 0.2$. } 
\label{fig:active_passive_cycle_doamin_size}
\end{figure}

{\it Results.--}  
In order to understand how microswimmers affect the phase separation dynamics of binary fluids, we  simulate a collection of active swimmers, with volume fraction $\phi$, in symmetric 50:50 mixtures of phase separating Newtonian fluids (Fig.~\ref{fig:SNapshots_pure_passive_active_comparison}).  In the absence of the swimmers, the fluid domains undergo a coarsening process until full phase separation is attained from an initially fully mixed state (Fig.~\ref{fig:SNapshots_pure_passive_active_comparison}a). This changes dramatically when microswimmers are included. Starting from a fully mixed state, the coarsening is arrested before the full phase separation (Fig.~\ref{fig:SNapshots_pure_passive_active_comparison}b). Interestingly, when starting from a fully phase separated state, a partly mixed state similar to the previous case is recovered (Fig.~\ref{fig:SNapshots_pure_passive_active_comparison}c). The final arrested state corresponds to a bicontinuous network of the fluids separated by an interface (Fig.~\ref{fig:SNapshots_pure_passive_active_comparison}d). The arrested coarsening corresponding to the appearance of a typical domain size $L_{\mathrm{steady}} << L_{\mathrm{system}}$ of the percolating network (Fig~\ref{fig:SNapshots_pure_passive_active_comparison}e). These observations suggest, that the hydrodynamic stresses created by the swimmers can mix the two fluids, leading to a dynamic foam-like state. 

To study the coarsening dynamics in more detail, we calculate the time-evolution of the domain length scale $L(t)$ (Fig.~\ref{fig:domain_pure_passive_active_comparison}).
In the absence of the active swimmers, the pure fluid undergoes spinodal decomposition, and the growth of the domain length-scale follows $L(t)\propto t$ (black symbols in Fig.~\ref{fig:domain_pure_passive_active_comparison}b), in agreement with previous studies~\cite{Kendon_JFM_2001,Tiribocchi_PRL_2015}. 
When microswimmers are added, $L(t)$ is observed to have a plateau for both particle volume fractions considered ($\phi\approx 5\%$ and $20\%$), independently of the initial conditions. Both the fully mixed starting configuration as well as the fully phase separated state, lead to approximately similar $L(t)$ in the steady state (see open and full symbols in Fig.~\ref{fig:domain_pure_passive_active_comparison}a and b). Interestingly, the foam-like states are observed for both extensile (pusher) and contractile (puller) swimmers.
We hypothesize, that the phenomenon arises from active pumping by the swimmers at, or near, the interfaces, similarly to non-motile point-like force dipoles in continuum fluid-fluid systems~\cite{Tiribocchi_PRL_2015, gulati2026bicontinuity}.
However, in our case the swimmers have a finite size and are self-propelled.

To understand the role of hydrodynamic stresses in detail, we first consider neutral swimmers which self-propel but do not carry force dipoles  $B_1>0$; $B_2=0$ (Fig.~\ref{fig:neutral_CaB1_0p1}).  
We do not observe arrested phase separation with neutral swimmers. The domain length-scale $L(t)$ increases until full phase separation is reached (Fig.~\ref{fig:neutral_CaB1_0p1}b). Similarly, starting from a phase separated state, $L(t)\approx L_{\mathrm{system}}$. This can be understood by considering wetting induced trapping of the swimmers~\cite{mishra2026interface}. Above a critical $\mathrm{Ca}^*_{B_1}$ the swimmers are not trapped but  move through the interface~\cite{mishra2026interface} (see {\it e.g.} Fig.~\ref{fig:neutral_CaB1_0p1}a for $\mathrm{Ca}_{B_1}\approx 0.1$) and no mixing of the two fluids is observed. When the $u_0$ is reduced below the $\mathrm{Ca}^*_{B_1}$ , the swimmers are permanently trapped at the interface, and similar behaviour to passive colloidal particles is recovered (see {\it e.g.} red and magenta symbols in Fig.~\ref{fig:neutral_CaB1_0p1}b).

In a general case, the swimmers both self-propel and carry a force dipole ($B_1 >0;~B_2\neq 0$) (Fig.~\ref{fig:phase_diagram}). When $\mathrm{Ca}_{B_1}>\mathrm{Ca^*}_{B_1}\approx 0.05$ the swimmers are not permanently trapped but can cross the interface\cite{mishra2026interface} (Fig.~\ref{fig:neutral_CaB1_0p1}). This gives rise to an interaction timescale $\tau_{i}\approx R/u_0\sim 1/B_1$, which corresponds to the time it takes a particle (radius $R$) to cross an interface. During this time, the swimmer force dipoles can stretch the interface. The stretching is resisted by the interfacial tension $\gamma$, leading to a stretching rate $\tau^{-1}_s\sim B_2/\gamma$, similarly to what is proposed in~\cite{gulati2026bicontinuity}. For a constant interfacial tension $\gamma$, this suggests that the critical pumping $B^*_2$ rate giving the transition between full phase separation and the observed bicontinuous state, has a linear form $B_2^*\approx B_1+ C$, where the constant $C$ corresponds non-motile (shaker) swimmers. 
The prediction agrees reasonably well with our observations (solid lines in Fig.~\ref{fig:phase_diagram}). A linear $L\approx AB_1+C$ fit in $\phi\approx 20\%$ sample,  gives 
$A\approx 1.3\pm 0.1$
(Fig~\ref{fig:phase_diagram}b), while for $\phi\approx 5\%$,  
$A\approx 2.6 \pm 0.1$
is observed (Fig.~\ref{fig:phase_diagram}a). This discrepancy might arise from the reduced number of swimmers in the system, which renders the simple scaling argument less accurate. Both volume fractions predict very similar behaviour for shakers (small $\mathrm{Ca}_{B_1}$ values in Fig.~\ref{fig:phase_diagram}).

Interestingly, the bicontinuous states are observed for both, extensile (pusher) and contractile (puller), swimmers (Fig.~\ref{fig:domain_pure_passive_active_comparison} and Fig.~\ref{fig:phase_diagram}). This can be understood by considering the reorientation of force dipoles at fluid-fluid interfaces~\cite{Gidituri_PRF_2022}. The pusher (puller) swimmers align along (perpendicular to) an interface~\cite{Gidituri_PRF_2022}. Due to symmetry, this leads to  stretching of the interface in both cases.

The foam-like state is marked by a characteristic  length-scale $L << L_{\mathrm{system}}$ (see {\it e.g.} late times in Fig.~\ref{fig:domain_pure_passive_active_comparison}). The steady state $L$ arises from the competition between stabilising interfacial forces and active mixing, whose strength depends from both the force dipole strength $B_2$ and the swimmer volume fraction $\phi$. In the emulsion state, $L$ is observed to decrease with increasing $B_2$ and $\phi$ (see intermediate values in Fig.~\ref{fig:B2_and_Phi_effect}a and b). At high $B_2$ and $\phi$ saturation of $L$ is observed (Fig.~\ref{fig:B2_and_Phi_effect}), similarly to continuum simulations~\cite{Tiribocchi_PRL_2015}. 
At the intermediate $B_2$ and $\phi$, our data supports scaling $L\sim B_2^{-1.7\pm 0.3}$, and $L\sim \phi^{-1.5\pm 0.2}$. Assuming a uniform swimmer distribution, the $B_2$ and $\phi$ can be mapped into a continuum activity $\zeta\sim B_2\phi^{2/3}$ using the strength of the force dipole flow $u\sim B_2/r^2$ with a typical particle separation $l\sim \phi^{-1/3}$. This gives $L\sim\zeta^{-1.7\ldots -1}$, which agrees  well with $L\sim \zeta^{-1.36}$ predicted for bicontinuous states in passive/active fluid mixtures~\cite{gulati2026bicontinuity}.

These observations demonstrate that the swimmer activity can be used to control the morphology of the steady state. In the extreme case, one can turn off the activity in the foam state (Fig.~\ref{fig:active_passive_cycle_doamin_size}). 
The domain size $L(t)$ starts to grow until kinetic arrest is observed (green symbols in Fig.~\ref{fig:active_passive_cycle_doamin_size}a).  This state corresponds to a bijel-like state where the passive particles stabilise the interface between the two fluids (Fig.~\ref{fig:active_passive_cycle_doamin_size}d) akin to what is observed with passive colloids ~\cite{Stratford_Science_2005}. The pair correlation function $g(r)$ is observed to be similar in the kinetically arrested state, for both active and passive quench protocols (blue and green curves in Fig.~\ref{fig:active_passive_cycle_doamin_size}b). During the active mixing state, the $g(r)$ is more liquid like (yellow curve in Fig.~\ref{fig:active_passive_cycle_doamin_size}c) as can be expected due to the swimming of the particles. This provides an active route to realise bijels, where the formation  can be driven periodically by alternating between active and passive cycles (Fig.~\ref{fig:active_passive_cycle_doamin_size}a). This could be realised experimentally by light sensitive bacteria by alternating  shading on the sample~\cite{arlt2018painting}.

{\it Summary.--} Using hydrodynamic simulations, we have investigated the phase morphology of binary fluids in the presence of finite size  microswimmers. Our results reveal a spontaneous emulsification by the swimmers, where the two fluids organise into a bicontinous foam-like state. The results are rationalised by considering two active components: source dipole (self-propulsion) and force dipole (mixing). We show that increased interfacial area is stabilised by the hydrodynamic pumping arising from the force dipole flows, in agreement with continuum models~\cite{Tiribocchi_PRL_2015,Fernando_PRL_2022}. Our system  provides a microscopic realisation of bicontinuous states observed in active/passive binary mixtures~\cite{gulati2026bicontinuity}, by considering self-propelling swimmers with a finite size. We show that when the self-propulsion is increased beyond a threshold, the foam-like state is destabilised and full phase separation is observed. These predictions can be realised experimentally by extending bacterial experiments in asymmetric phase separating binary fluid mixtures~\cite{cheon2025motility,Yang_Arxiv_2025,Chang_Arxiv_2025,Yang_Arxiv_2026,Wang_bioarxiv_2026} to (close to) symmetric mixtures. The mapping to active capillary numbers allows a straight forward path to experimental verification of our predictions. 

\begin{acknowledgments}
We thank A. W\"urger and A. Aubret for insightful discussions.
HG, SS and JSL acknowledge IdEx (Initiative d'Excellence) Bordeaux, the  French  National  Research  Agency  through  Contract No. ANR-19-CE06-0012 for funding, and Curta cluster for computational time.
\end{acknowledgments}

\bibliography{apssamp}

@PREAMBLE{
 "\providecommand{\noopsort}[1]{}" 
 # "\providecommand{\singleletter}[1]{#1}%" 
}

@ARTICLE{Kendon_JFM_2001,
   author       = "JViven M. Kendon and Michael E. Cates and Ignacio Pagonabarraga and J.C. Desplat",
   title        = "Inertial effects in three dimensional spinodal
decomposition of a symmetric binary fluid
mixture: A lattice Boltzmann study",
   journal      = "Journal of Fluid Mechanics",
   volume       = "440", 
   pages        = "147-203",
   year         = "2001",
}

@article{Perlekar_PRL_2014,
  title = {Spinodal Decomposition in Homogeneous and Isotropic Turbulence},
  author = {Perlekar, Prasad and Benzi, Roberto and Clercx, Herman J. H. and Nelson, David R. and Toschi, Federico},
  journal = {Phys. Rev. Lett.},
  volume = {112},
  issue = {1},
  pages = {014502},
  numpages = {5},
  year = {2014},
}

@article{Perlekar_POF_2012,
author = {Perlekar,Prasad  and Biferale,Luca  and Sbragaglia,Mauro  and Srivastava,Sudhir  and Toschi,Federico },
title = {Droplet size distribution in homogeneous isotropic turbulence},
journal = {Physics of Fluids},
volume = {24},
number = {6},
pages = {065101},
year = {2012},

}

@article{Dennis_PRL_2021,
  title = {Catastrophic Phase Inversion in High-Reynolds-Number Turbulent Taylor-Couette Flow},
  author = {Bakhuis, Dennis and Ezeta, Rodrigo and Bullee, Pim A. and Marin, Alvaro and Lohse, Detlef and Sun, Chao and Huisman, Sander G.},
  journal = {Phys. Rev. Lett.},
  volume = {126},
  issue = {6},
  pages = {064501},
  numpages = {6},
  year = {2021},
}

@ARTICLE{Perlekar_SR_2017,
   author       = "Prasad, Perlekar and Nairita, Pal and Rahul, Pandit",
   title        = "Two-dimensional Turbulence in
Symmetric Binary-Fluid Mixtures:
Coarsening Arrest by the Inverse
Cascade",
   journal      = "Scientific Reports",
   volume       = "7", 
   pages        = "44589",
   year         = "2017",
}

@ARTICLE{Cates_SM_2008,
   author       = "Cates, Michael E. and Clegg, Paul S.",
   title        = "Bijels: a new class of soft materials",
   journal      = "Soft Matter",
   volume       = "4", 
   pages        = "2132-2138",
   year         = "2008",
}

@ARTICLE{Stratford_Science_2005,
   author       = "Stratford, K. and Adhikari, R. and Pagonabarraga, I. and Desplat, J.-C. and Cates, M. E.",
   title        = "Colloidal Jamming at Interfaces: A Route to Fluid-Bicontinuous Gels",
   journal      = "Science",
   volume       = "309", 
   pages        = "2198--2201",
   year         = "2005",
}

@article{Hari_PoF_2021,
author = {Gidituri,Harinadha  and Würger,Alois  and Stratford,Kevin  and Lintuvuori,Juho S. },
title = {Dynamics of a spherical colloid at a liquid interface: A lattice Boltzmann study},
journal = {Physics of Fluids},
volume = {33},
number = {5},
pages = {052110},
year = {2021},

}

@Article{Juho_SM_2016,
author ="Lintuvuori, Juho S. and Brown, Aidan T. and Stratford, Kevin and Marenduzzo, Davide",
title  ="Hydrodynamic oscillations and variable swimming speed in squirmers close to repulsive walls",
journal  ="Soft Matter",
year  ="2016",
volume  ="12",
issue  ="38",
pages  ="7959-7968",
}

@Article{Zaiyi_EPJE_2018,
author ="Zaiyi Shen and Alois Würger and Juho S. Lintuvuori",
title  ="Hydrodynamic interaction of a self-propelling particle with a wall",
journal  ="The European Physical Journal E",
year  ="2018",
volume  ="41",
issue  ="39",
pages  ="39",
}

@article{magar2003nutrient,
  title={Nutrient Uptake by a Self-Propelled Steady Squirmer},
  author={Magar, Vanesa and Goto, Tomonobu and Pedley, Timothy J},
  journal={The Quarterly Journal of Mechanics and Applied Mathematics},
  volume={56},
  number={1},
  pages={65--91},
  year={2003},
  publisher={Oxford University Press}
}

@ARTICLE{Pagonabarraga_JNFM_2010,
   author       = "I. Llopis and I. Pagonabarraga",
   title        = "Hydrodynamic interactions in squirmer motion: Swimming with a neighbour and close to a wall",
   journal      = "Journal of Non-Newtonian Fluid Mechanics",
   volume       = "165", 
   pages        = "946-952",
   year         = "2010",
}

@article{Lin_JFM_2011, 
title={Stirring by squirmers},
  author={Lin, Zhi and Thiffeault, Jean-Luc and Childress, Stephen},
  journal={Journal of Fluid Mechanics},
  volume={669},
  pages={167--177},
  year={2011},
  publisher={Cambridge University Press}
}

@article{Fernando_PRL_2022,
  title = {Activity-Suppressed Phase Separation},
  author = {Caballero, Fernando and Marchetti, M. Cristina},
  journal = {Phys. Rev. Lett.},
  volume = {129},
  issue = {26},
  pages = {268002},
  numpages = {6},
  year = {2022},
  month = {Dec},
  publisher = {American Physical Society},
  doi = {10.1103/PhysRevLett.129.268002},
  url = {https://link.aps.org/doi/10.1103/PhysRevLett.129.268002}
}

@article{Adkins_Science_2022,
  title={Dynamics of active liquid interfaces},
  author={Adkins, Raymond and Kolvin, Itamar and You, Zhihong and Witthaus, Sven and Marchetti, M Cristina and Dogic, Zvonimir},
  journal={Science},
  volume={377},
  number={6607},
  pages={768--772},
  year={2022},
  publisher={American Association for the Advancement of Science}
}

@article{Tayar_NatMat_2023,
  title={Controlling liquid--liquid phase behaviour with an active fluid},
  author={Tayar, Alexandra M and Caballero, Fernando and Anderberg, Trevor and Saleh, Omar A and Cristina Marchetti, M and Dogic, Zvonimir},
  journal={Nature Materials},
  volume={22},
  number={11},
  pages={1401--1408},
  year={2023},
  publisher={Nature Publishing Group UK London}
}

@article{Tiribocchi_PRL_2015,
  title = {Active Model H: Scalar Active Matter in a Momentum-Conserving Fluid},
  author = {Tiribocchi, Adriano and Wittkowski, Raphael and Marenduzzo, Davide and Cates, Michael E.},
  journal = {Phys. Rev. Lett.},
  volume = {115},
  issue = {18},
  pages = {188302},
  numpages = {5},
  year = {2015},
  month = {Oct},
  publisher = {American Physical Society},
  doi = {10.1103/PhysRevLett.115.188302},
  url = {https://link.aps.org/doi/10.1103/PhysRevLett.115.188302}
}

@article{Ali_TB_2025,
  title={Oil--water interfaces of Pickering emulsions: microhabitats for living cell biocatalysis},
  author={Ali, Daniel Chikere and Pan, Tao and Wu, Qingping and Wang, Zhilong},
  journal={Trends in Biotechnology},
  year={2025},
  publisher={Elsevier}
}

@article{Gidituri_PRF_2022,
  title={Reorientation dynamics of microswimmers at fluid-fluid interfaces},
  author={Gidituri, Harinadha and Shen, Zaiyi and W{\"u}rger, Alois and Lintuvuori, Juho S},
  journal={Physical Review Fluids},
  volume={7},
  number={4},
  pages={L042001},
  year={2022},
  publisher={APS}
}

@article{Prasad_Science_2023,
  title={Alcanivorax borkumensis biofilms enhance oil degradation by interfacial tubulation},
  author={Prasad, M and Obana, N and Lin, S-Z and Zhao, S and Sakai, K and Blanch-Mercader, C and Prost, J and Nomura, N and Rupprecht, J-F and Fattaccioli, J and others},
  journal={Science},
  volume={381},
  number={6659},
  pages={748--753},
  year={2023},
  publisher={American Association for the Advancement of Science}
}

@article{Yang_Arxiv_2025,
  title={Active motility and wetting cooperatively regulate liquid-liquid phase separation},
  author={Yang, Dixi and Wang, Anheng and Wang, Chunming and Tanaka, Hajime and Yuan, Jiaxing},
  journal={arXiv preprint arXiv:2511.18077},
  year={2025}
}

@article{Chang_Arxiv_2025,
  title={Collective bacterial motion drives interfacial waves and shape dynamics in phase-separated droplets},
  author={Chang, Kan and Li, Yulin and Yuan, Ming and Sano, Masaki and You, Zhihong and Zhang, HP},
  journal={arXiv preprint arXiv:2511.12621},
  year={2025}
}

@misc{gulati2026bicontinuity,
      title={Bicontinuity in active phase separation}, 
      author={Paarth Gulati and Liang Zhao and Michio Tateno and Omar A. Saleh and Zvonimir Dogic and M. Cristina Marchetti},
      year={2026},
      eprint={2601.03221},
      archivePrefix={arXiv},
      primaryClass={cond-mat.soft},
      url={https://arxiv.org/abs/2601.03221}, 
}

@article{mishra2026interface,
  title={Interface crossing behavior of prolate microswimmers: Thermo and hydrodynamics},
  author={Mishra, Rishish and Pothukuchi, Harish and Gidituri, Harinadha and Lintuvuori, Juho},
  journal={Physical Review Fluids},
  volume={11},
  number={1},
  pages={014002},
  year={2026},
  publisher={APS}
}

@article{cates2025active,
  title={Active phase separation: new phenomenology from non-equilibrium physics},
  author={Cates, Michael E and Nardini, Cesare},
  journal={Reports on Progress in Physics},
  volume={88},
  number={5},
  pages={056601},
  year={2025},
  publisher={IOP Publishing}
}

@article{Yang_Arxiv_2026,
  title={Wetting-coupled phase separation as an energetic mechanism for active bacterial adhesion},
  author={Yang, Dixi and Wang, Anheng and Huang, Jia and Zhuo, Xiaofeng and Wang, Chunming and Tanaka, Hajime and Yuan, Jiaxing},
  journal={arXiv preprint arXiv:2601.06754},
  year={2026}
}

@article{Wang_bioarxiv_2026,
  title={Wetting-mediated extracellular phase separation drives long-range cell adhesion},
  author={Wang, Anheng and Yang, Dixi and Zhang, Haijiao and Paunov, Vesselin and Tian, Shuo and Dong, Lei and Tanaka, Hajime and Yuan, Jiaxing and Wang, Chunming},
  journal={bioRxiv},
  pages={2026--01},
  year={2026},
  publisher={Cold Spring Harbor Laboratory}
}

@article{arlt2018painting,
  title={Painting with light-powered bacteria},
  author={Arlt, Jochen and Martinez, Vincent A and Dawson, Angela and Pilizota, Teuta and Poon, Wilson CK},
  journal={Nature communications},
  volume={9},
  number={1},
  pages={768},
  year={2018},
  publisher={Nature Publishing Group UK London}
}

@article{diaz2025activity,
  title={Activity-driven emulsification of phase-separating binary mixtures},
  author={D{\'\i}az, Javier and Pagonabarraga, Ignacio},
  journal={Physical Review Letters},
  volume={134},
  number={9},
  pages={098301},
  year={2025},
  publisher={APS}
}

@article{singh2019hydrodynamically,
  title={Hydrodynamically interrupted droplet growth in scalar active matter},
  author={Singh, Rajesh and Cates, ME},
  journal={Physical review letters},
  volume={123},
  number={14},
  pages={148005},
  year={2019},
  publisher={APS}
}

@article{cheon2025motility,
  title={Motility modulates the partitioning of bacteria in aqueous two-phase systems},
  author={Cheon, Jiyong and Choi, Kyu Hwan and Modica, Kevin J and Mitchell, Robert J and Takatori, Sho C and Jeong, Joonwoo},
  journal={Physical Review Letters},
  volume={135},
  number={12},
  pages={128401},
  year={2025},
  publisher={APS}
}

@article{cheon2024motile,
  title={Motile bacteria crossing liquid--liquid interfaces of an aqueous isotropic--nematic coexistence phase},
  author={Cheon, Jiyong and Son, Joowang and Lim, Sungbin and Jeong, Yundon and Park, Jung-Hoon and Mitchell, Robert J and Kim, Jaeup U and Jeong, Joonwoo},
  journal={Soft Matter},
  volume={20},
  number={36},
  pages={7313--7320},
  year={2024},
  publisher={Royal Society of Chemistry}
}

@software{kevinstratford_2024_12822477,
  author       = {kevinstratford and
                  Oliver Henrich and
                  jlintuvuori and
                  dmarendu and
                  qikaifzj and
                  austin1997 and
                  shanCHEN123 and
                  ludwig-cf and
                  jurijsab and
                  Sergio Granados Leyva and
                  sumeshpt},
  title        = {ludwig-cf/ludwig: Ludwig 0.22.0},
  month        = jul,
  year         = 2024,
  publisher    = {Zenodo},
  version      = {ludwig-0.21.0},
  doi          = {10.5281/zenodo.12822477},
  url          = {https://doi.org/10.5281/zenodo.12822477},
}

@article{ludwigcode,
    journal = {Ludwig: A lattice Boltzmann code for complex fluids, https://github.com/ludwig-cf/ludwig}
}

\end{document}